\documentclass[aps,prd,
superscriptaddress,
showkeys,
showpacs]{revtex4}

\usepackage[english]{babel}
\usepackage{amsmath,bm}
\usepackage{amssymb}
\usepackage{enumitem}
\usepackage{textcomp}
\usepackage{graphicx}
\usepackage{dsfont}
\usepackage{color}
\usepackage{hyperref}
\usepackage{mathrsfs}
\graphicspath{{Figures/}}

\definecolor{purple}{rgb}{0.8,0,0.6}

\begin{document}

\title{Kinetic approach to the Schwinger effect during inflation}

\author{E.V.~Gorbar}
\affiliation{Physics Faculty, Taras Shevchenko National University of Kyiv, 64/13, Volodymyrska Str., 01601 Kyiv, Ukraine}
\affiliation{Bogolyubov Institute for Theoretical Physics, 14-b, Metrologichna Str., 03680 Kyiv, Ukraine}

\author{A.I.~Momot}
\affiliation{Physics Faculty, Taras Shevchenko National University of Kyiv, 64/13, Volodymyrska Str., 01601 Kyiv, Ukraine} 

\author{O.O.~Sobol}
\affiliation{Institute of Physics, Laboratory for Particle Physics and Cosmology (LPPC), \'{E}cole Polytechnique F\'{e}d\'{e}rale de Lausanne (EPFL), CH-1015 Lausanne, Switzerland}
\affiliation{Physics Faculty, Taras Shevchenko National University of Kyiv, 64/13, Volodymyrska Str., 01601 Kyiv, Ukraine}
\email{oleksandr.sobol@epfl.ch}

\author{S.I.~Vilchinskii}
\affiliation{Institute of Physics, Laboratory for Particle Physics and Cosmology (LPPC), \'{E}cole Polytechnique F\'{e}d\'{e}rale de Lausanne (EPFL), CH-1015 Lausanne, Switzerland}
\affiliation{D\'{e}partement de Physique Th\'{e}orique, Center for Astroparticle Physics, Universit\'{e} de Gen\`{e}ve, 1211 Gen\`{e}ve 4, Switzerland}
\affiliation{Physics Faculty, Taras Shevchenko National University of Kyiv, 64/13, Volodymyrska Str., 01601 Kyiv, Ukraine}

\date{\today}
\pacs{04.62.+v, 47.75.+f, 98.80.Cq, 98.62.En}

\begin{abstract}
Using the kinetic approach, we study the impact of the charged particle dynamics due to the Schwinger effect on the electric field evolution during inflation. As a simple model of the electric field generation, we consider the kinetic coupling of the electromagnetic field to the inflaton via the term $f^{2}(\phi)F_{\mu\nu}F^{\mu\nu}$ with the Ratra coupling function $f=\exp(\beta\phi/M_{p})$. The production of charged particles is taken into account in the Boltzmann kinetic equation through the Schwinger source term. Produced particles are thermalized due to collisions which we model by using the collision integral in the self-consistent relaxation time approximation.
We found that the current of created particles exhibits a non-Markovian character and cannot be described by a simple Ohm's law relation $j\propto E$. On the contrary, the electric current as well as the electric field are oscillatory functions of time with decreasing amplitudes and a phase difference due to the ballistic motion of charged carriers.
Our qualitative results are checked by using a hydrodynamic approach. Deriving a closed system of equations for the number, current, and energy densities of charged particles
and determining its solution, we find a good agreement with the results obtained in the kinetic approach.
\end{abstract}

\keywords{Schwinger effect, Boltzmann kinetic equation, hydrodynamic approach}

\maketitle

\section{Introduction}
\label{sec-intro}

The detection of magnetic fields in cosmic voids through the gamma-ray observations of distant blazars
\cite{Neronov:2010,Tavecchio:2010,Taylor:2011,Dermer:2011,Caprini:2015} is a remarkable discovery. Combining these observations with the data 
of the cosmic microwave background (CMB) \cite{Planck:2015-pmf,Sutton:2017,Jedamzik:2018,Paoletti:2018,Giovannini:2018b} and ultra-high-energy 
cosmic rays \cite{Bray:2018} observations constrains the strength of these fields to $10^{-18}\lesssim B\lesssim 10^{-9}\,$G. The extremely
large coherence scale $\lambda_{B}\gtrsim 1\,$Mpc strongly suggests the cosmological origin of magnetic fields observed in voids.
Thus, the study of such magnetic fields and their physical characteristics may shed light on the earliest stages of the Universe's history. Although phase transitions in the early Universe could generate primordial magnetic fields of the necessary strength \cite{Hogan:1983,Quashnock:1989,Vachaspati:1991,Cheng:1994,Sigl:1997,Ahonen:1998}, the coherence length of such fields
is determined by the horizon size at the moment of a phase transition and is much less than Mpc today. This leaves inflation and/or reheating as the only natural mechanism that could generate magnetic fields on such a large scale \cite{Turner:1988,Ratra:1992}. 

As is well known, one of the key requirements necessary for the generation of magnetic fields during inflation is the breaking of the 
conformal symmetry in the electromagnetic (EM) sector \cite{Parker:1968}. This could be done by introducing the 
interaction of the EM field with a scalar or pseudoscalar inflaton field or with the curvature scalar (see, the pioneering works 
\cite{Turner:1988,Ratra:1992,Garretson:1992,Dolgov:1993}). In addition to magnetic fields, typically, strong electric fields are 
also generated. In fact, electric fields are minimally of the same order as magnetic fields in the axial coupling model 
\cite{Durrer:2011,Fujita:2015,Figueroa:2018,Sobol:2019}, while they are much stronger 
than the magnetic fields in the kinetic coupling model with the coupling function decreasing in time \cite{Martin:2008,Demozzi:2009,Kanno:2009,Sobol:2018} (e.g., in the Ratra model considered in this 
work). Obviously, such strong electric fields cannot survive during the 
postinflationary epoch because they quickly dissipate in the highly conducting medium of the primordial plasma. Nevertheless, 
electric fields might be still very important during inflation and preheating as they can significantly affect the magnetic field 
evolution \cite{Kobayashi:2019} and produce the charged particles due to the Schwinger effect.

In view of the Schwinger effect \cite{Sauter:1931,Heisenberg:1936,Schwinger:1951}, a strong electric field $E$ creates pairs of 
charged particles  whose production rate is $\propto E^2\exp(-\pi m^2c^3/|eE|\hbar)$ in the Minkowski spacetime. Obviously, this 
rate is exponentially suppressed unless the electric field exceeds
the critical value $E_{\rm cr}=m^2_ec^3/(e\hbar) \simeq 1.3\times 10^{16}\,{\rm V/cm}$ for the lightest charged particles, which are 
the electrons and positrons in the Standard Model. Although predicted very long ago, the Schwinger pair production was never observed in the laboratory because the critical field value is extremely large.
Even the new generation of laser systems such as the Extreme Light Infrastructure  \cite{Dunne:2008,Heinzl:2008} and the Exawatt Center for Extreme Light Studies \cite{Bashinov:2014} could produce the peak field strengths of only $\sim 10^{-2}E_{\rm cr}$ which would still provide a very small production rate. The Schwinger effect may occur in other physical systems such as relativistic heavy-ion collisions
\cite{Casher:1979,Glendenning:1983,Kajantie:1985,Gatoff:1987,Blaschke:2019}, decays of ``hadronic strings'' during the process of 
hadronization \cite{Andersson:1983}, condensed matter physics \cite{Zener:1934}, and, finally, in the early Universe \cite{Kobayashi:2014, Froeb:2014, Bavarsad:2016, Stahl:2016a, Stahl:2016b, Hayashinaka:2016a,  Hayashinaka:2016b,  Sharma:2017, Tangarife:2017, Bavarsad:2018, Hayashinaka:2018, Hayashinaka:thesis, Stahl:2018, Geng:2018, Giovannini:2018a, Kitamoto:2018,Banyeres:2018,Chua:2019,Shakeri:2019} which is of the prime interest for our investigation.

In an expanding universe, analytic expressions for the Schwinger pair production rate or the current of produced particles can be analytically obtained only in the simplest case of a constant electric field in de Sitter spacetime \cite{Kobayashi:2014,  Froeb:2014,  Bavarsad:2016, Stahl:2016a, Stahl:2016b, Hayashinaka:2016a,  Hayashinaka:2016b,  Sharma:2017, Tangarife:2017, Bavarsad:2018, Hayashinaka:2018, Hayashinaka:thesis, Stahl:2018}.  However, as shown in Ref.\cite{Giovannini:2018a}, this case is unphysical since maintaining a constant electric field in an exponentially expanding universe would require the existence of \textit{ad hoc} currents that could violate the second law of thermodynamics.  Various cases of (1+1)-dimensional \cite{Froeb:2014,Stahl:2016a,Stahl:2016b}, (2+1)-dimensional \cite{Bavarsad:2016}, and (3+1)-dimensional \cite{Kobayashi:2014,Hayashinaka:2016a,Hayashinaka:2016b,Hayashinaka:2018,Hayashinaka:thesis} de Sitter spacetime with scalar \cite{Froeb:2014,Kobayashi:2014,Bavarsad:2016,Hayashinaka:2016b,Hayashinaka:2018,Hayashinaka:thesis,Stahl:2018} and spinor charged fields \cite{Stahl:2016a,Stahl:2016b,Hayashinaka:2016a,Hayashinaka:2018,Hayashinaka:thesis}, including also an external magnetic field \cite{Bavarsad:2018}, were investigated.  Unfortunately, the current of created particles obtained by direct averaging of the corresponding current operator contains ultraviolet divergences, which can be regularized using the adiabatic subtraction \cite{Kobayashi:2014,Hayashinaka:2016a} or point-splitting method \cite{Hayashinaka:2016b}. Although these techniques remove the divergent parts, the finite part is not uniquely defined and depends on the applied subtraction scheme. It should be noted that the cosmological Schwinger effect contains interesting features which are absent in its flat-space counterpart and are rather counterintuitive such as (i) the infrared hyperconductivity for a scalar field when the conductivity becomes very large in the limit of small mass and (ii) the negative conductivity in the weak field regime $|eE|\ll H^{2}$ (here $H$ is the Hubble parameter) \cite{Hayashinaka:2016a,Hayashinaka:thesis}  which may enhance the electric field \cite{Stahl:2018}.  However, it was recently argued in Ref.~\cite{Banyeres:2018} that the reported negative conductivity is spurious and originates from the nonlinear corrections to the Maxwell action and the logarithmic running of the coupling constant.

An attempt to combine the generation of EM fields due to the kinetic coupling with the Schwinger effect and backreaction into a consistent picture was made in Ref.~\cite{Kitamoto:2018}, where it was shown that the expression for the Schwinger current in time-dependent electric background in the strong-field regime has the same functional dependence as in the case of a constant electric field. In recent works \cite{Sobol:2019,Sobol:2018}, it was shown that the Schwinger effect becomes relevant only on the last stages of inflation when the inflaton is close to the minimum of its potential (see also Ref.~\cite{Shakeri:2019}). It was found that the Schwinger effect abruptly decreases the value of the electric field helping to finish the inflation stage and enter the stage of preheating.  Since charged particles produced by the Schwinger effect can leave their unique imprints on the power spectra of primordial perturbations \cite{Chua:2019}, the latter can be used in order to testify for the existence of the primordial electric fields.

 Although the cosmological Schwinger effect has been intensively studied in the last decades, still there are some unsolved problems which should be addressed. First of all, the use of the expressions for the Schwinger current in a constant electric field  is justified only if the electric field evolves adiabatically, i.e.,  when the characteristic timescale of its variation is much larger than the Hubble time. In the opposite case of a quickly varying electric field, the charge carriers cannot change their direction of motion  fast enough with the change  of the electric field because of their inertial properties. In addition, it is also necessary to take into account the process of thermalization of the produced particles due to their scattering. Therefore, the dynamical aspects of the produced charged particles could play a potentially very important role in the investigation of the Schwinger effect and the evolution of the electric field during inflation. This provides the main motivation to apply the kinetic approach in order to study the Schwinger effect during inflation in the present paper.

The Boltzmann kinetic equation with a Schwinger source term which describes the pair creation was successfully employed in a time-varying electric field in flat spacetime \cite{Kajantie:1985,Gatoff:1987,Blaschke:2019,Kluger:1991,Kluger:1992,Schmidt:1998,Kluger:1998,Bloch:1999,Alkofer:2001}. The explicit form of the source term can be derived  by using the field-theoretic methods \cite{Rau:1994,Schmidt:1998,Kluger:1998} or modeled by an expression which has the correct asymptotic behavior \cite{Kajantie:1985,Gatoff:1987,Kluger:1991,Kluger:1992}. In our work, we  extend these studies to the case of inflation-produced electric fields in an expanding universe. Our results will be also checked within the corresponding hydrodynamic approach  which is much less sensitive to
the explicit form of the Schwinger source term.

In order to describe the generation of electromagnetic fields during inflation, for the purpose of this paper, it is sufficient to
consider a simple case of the kinetic coupling of the EM field to the inflaton field via the term $f^2(\phi)F_{\mu\nu}F^{\mu\nu}$ introduced 
by Ratra \cite{Ratra:1992} and revisited many times in the
literature \cite{Giovannini:2001,Bamba:2004,Martin:2008,Demozzi:2009,Kanno:2009,Ferreira:2013,Ferreira:2014,Vilchinskii:2017,Sharma:2017b,Shtanov:2018}. The Ratra coupling function $f=\exp(\beta\phi/M_{p})$ is a decreasing function of time  that does not cause a strong coupling problem during inflation and leads to the production of very strong electric field \cite{Martin:2008,Demozzi:2009,Kanno:2009,Sobol:2018}. For the sake of definiteness, the numerical results are obtained in the Starobinsky model of inflation \cite{Starobinsky:1980} which is one of the models most favored by the CMB observations \cite{Planck:2018inf}.  The kinetic approach and the results obtained in this paper can be easily generalized to other inflationary models.

This paper is organized as follows. We set up our model in Sec.~\ref{sec-model}. The Boltzmann equation is considered in Sec.~\ref{sec-kinetic} and the hydrodynamic approach is developed in Sec.~\ref{sec-hydro}. Our numerical results are presented in Sec.~\ref{sec-numerical} and the summary is given in Sec.~\ref{sec-concl}.  In what follows we use the natural units and set $\hbar=c=1$.

\section{Kinetic coupling model}
\label{sec-model}
\vspace{5mm}

 As  stated in the Introduction, we consider in this paper a generic model with a single scalar inflaton field $\phi$ kinetically 
coupled to the EM field, and a charged field $\chi$ whose quanta are produced by a strong electric field. The corresponding action has the 
form
\begin{equation}
\label{action}
S=\int d^{4}x \sqrt{-{\rm det}(g_{\mu\nu})}\left[\frac{1}{2}g^{\mu\nu}\partial_{\mu}\phi\partial_{\nu}\phi-V(\phi)-\frac{1}{4}f^{2}(\phi)F_{\mu\nu}F^{\mu\nu}-e j^{\mu}_{\chi}A_{\mu}\right],
\end{equation}
where $g_{\mu\nu}$ is the metric tensor, $F_{\mu\nu}=\partial_{\mu}A_{\nu}-\partial_{\nu}A_{\mu}$ is the EM field tensor, $V(\phi)$ is the inflaton potential, $f(\phi)$ is the kinetic coupling function, $e$ is the elementary charge defined as $e^{2}/(4\pi)=\alpha_{\rm EM}\simeq 1/137$, and $j^{\mu}_{\chi}$ is the electric current of a charged field $\chi$ coupled to the gauge field $A_{\mu}$. This current is of principal interest for us in this paper. In order to find it, we describe 
the produced charged particles by their one-particle distribution function in the phase space and study its evolution using the Boltzmann kinetic equation.
For the sake of generality, we do not specify the field $\chi$ and derive all equations for a generic  case of boson or fermion charge carriers. For illustrative purposes, however, we will consider the case of an electron charged field which is the lightest particle in the Standard Model.

We would like to emphasize that the effective coupling constant which determines the interaction of the charged particles with the EM field 
in model (\ref{action}) is $e_{\rm eff}=e/f$ rather than $e$ because the EM kinetic term in action (\ref{action}) is not 
in a canonical form and contains additional factor $f^{2}$. Absorbing this extra factor by  redefining the EM potential
$A_{\mu}\to A_{\mu}/f$  rescales the charge $e\to e_{\rm eff}$. This implies that the intensity of the Schwinger pair production is tuned by the coupling function. 

The latter at the end of inflation when the inflaton field stops in the minimum of its potential should tend to unity in order to recover the correct value of the particle's charge.  In addition, the strong coupling problem during inflation is avoided if $e_{\rm eff}<1$ leading to the condition $f\gtrsim 1$ \cite{Demozzi:2009}. Finally,  for a monotonously decreasing coupling function during inflation, the electric component of the generated EM field dominates over the magnetic one \cite{Martin:2008,Kanno:2009,Sobol:2018}. Therefore, a good approximation is to neglect the influence of the magnetic field on the Schwinger effect that we will do in the rest of the paper.  

We  consider a spatially flat Friedmann-Lema\^{i}tre-Robertson-Walker universe whose metric tensor $g_{\mu\nu}=\mbox{diag}(1,-a^2,-a^2,-a^2)$ is determined by a scale factor $a(t)$. Its time evolution is described by the Friedmann equation
\begin{equation}
\label{eq-Friedmann}
H^{2}\equiv \left(\frac{\dot{a}}{a}\right)^{2}=\frac{1}{3M_{p}^{2}}\left(\rho_{\phi}+ \rho_{\rm EM}+\rho_{c}\right),
\end{equation}
where $M_{p}=(8\pi G)^{-1/2}=2.4\times 10^{18}\,{\rm GeV}$ is the reduced Planck mass, $\rho_{\phi}=\frac{1}{2}\dot{\phi}^{2}+V(\phi)$ is the energy density of the inflaton field, which  dominates during inflation, $\rho_{\rm EM}=\frac{f^{2}}{2}\left<E^{2}+B^{2}\right>$ is the energy density of the generated EM field, and $\rho_{c}$ is the energy density of the charged particles created due to the Schwinger effect.

Let us assume that the inflaton and electric fields are spatially homogeneous (of course, there are always some spatial inhomogeneities, 
however, they could be neglected in the first approximation). Varying
action (\ref{action}) with respect to the inflaton and EM fields, we  obtain the following Klein-Gordon-Fock and Maxwell 
equations:
\begin{eqnarray}
\label{KGF-2}
&&\ddot{\phi}+3H\dot{\phi}+\frac{dV}{d\phi}=f(\phi)f'(\phi)E^{2}(t),\\
\label{Maxwell-2}
&&\partial_{t}\left(a^{2}f^{2}E^{i}\right)=-a^{3}(t)ej^{i},
\end{eqnarray}
where $E^{i}=-a(t)F^{0i}$ and $j^{i}$ are the electric field and current three-vectors, respectively.
The classical electric field  which is present in Eq.~(\ref{Maxwell-2}) is generated from the quantum fluctuations through the quantum-to-classical transition when the corresponding EM modes cross the Hubble horizon and undergo amplification due to the interaction with the inflaton.

Equation~(\ref{Maxwell-2}) describes only the enhancement of modes, which have already crossed the Hubble horizon, and does not take into account the fact that the number of such modes grows as the inflation goes on. This can be fixed by introducing  a boundary term  that describes new modes crossing the Hubble horizon and undergoing the quantum-to-classical transition.  Such a term was derived in Ref.~\cite{Sobol:2018} by considering the equation for the evolution of the electric energy density $\rho_{E}=f^{2}E^{2}/2$. The corresponding equation reads as
\begin{equation}
\label{Maxwell-4}
\dot{\rho}_{E}+4H(t)\rho_{E}+2\frac{\dot{f}}{f}\rho_{E}=-a(t)e \left(\mathbf{E} \cdot \mathbf{j}\right) +\frac{H^{3}}{4\pi^{2}}\left[H^{2}+\left(\frac{\dot{f}}{f}\right)^{2}\right].
\end{equation}
Introducing a new variable $\mathcal{E}=eE=e_{\rm eff}\sqrt{2\rho_{E}}$, we obtain
\begin{equation}
\label{eq-for-electric-field}
\dot{\mathcal{E}}+2H(t)\mathcal{E}+2\frac{\dot{f}}{f}\mathcal{E}=-e_{\rm eff}^{2}\tilde{j}+e_{\rm eff}^{2}\frac{H^{3}}{4\pi^{2}\mathcal{E}}\left[H^{2}+\left(\frac{\dot{f}}{f}\right)^{2}\right],
\end{equation}
where $\tilde{j}=a(t)j$ is the  electric current measured by a comoving observer \cite{Subramanian:2016}.
The last term on the right-hand side  of the above equation is the sought boundary term. This term is relevant for a weak field $\mathcal{E}\lesssim e_{\rm eff}H^{2}$ when the contribution of new modes crossing the horizon is important. The Schwinger current  given by the first term on the right-hand side is negligible in this regime. If the electric field is strong $\mathcal{E}\gg e_{\rm eff}H^{2}$, then the boundary term can be neglected. Clearly, this term has to be excluded after the end of inflation when no new modes cross the Hubble horizon.

We would like to note  that $E_{\rm eff}=fE$ is the value of the electric field measured by a comoving observer. In fact, the electric energy density is expressed in terms of this effective electric field in the same way as in Minkowski space, i.e., $\rho_{E}=E_{\rm eff}^{2}/2$.  Since $\mathcal{E}=e_{\rm eff} E_{\rm eff}$ is expressed only through the effective charge and the effective electric field, it is a complete analogue of $eE$ in a flat spacetime. Therefore, we conclude that the Schwinger pair production rate is determined by $\mathcal{E}$ during inflation.  Thus, we have a system of equations governing the evolution of the scale factor, the inflaton and electric fields which contains the energy density $\rho_{c}$ and the current $\tilde{j}$ of the charged particles produced due to the Schwinger effect. These two quantities will be determined in the next  two sections using the kinetic and hydrodynamic approaches.

\section{Kinetic approach}
\label{sec-kinetic}
\vspace{5mm}

In this  section, we formulate the kinetic framework which describes the production and evolution of charged particles during inflation.

\subsection{Boltzmann equation}

For one-particle distribution function depending on contravariant coordinates and covariant momentum
$\mathcal{F}_{s}(x^{\mu},P_{j})$, the relativistic Boltzmann equation has the  simplest form  \cite{Weinberg-book,Gorbunov-book-v2}
\begin{equation}
\label{Boltzmann-covariant}
\left[P^{\mu}\frac{\partial}{\partial x^{\mu}}-e_{s}P^{\mu}F_{\mu j}\frac{\partial}{\partial P_{j}}\right] \mathcal{F}_{s}(x^{\nu},P_{k})=u^{\mu}P_{\mu}\left(\mathcal{S}[\mathcal{F}_{s}]+\mathcal{C}[\mathcal{F}_{s}]\right),
\end{equation}
where index $s=p,a$ corresponds to particles and antiparticles, respectively, $e_{p}=e$, $e_{a}=-e$, and $u^{\mu}$ is the local flow four-velocity.  As usual, the greek letters $\mu,\,\nu\,\ldots$ denote the Lorentz indices varying from 0 to 3 while the latin letters $i,\,j,\,k,\,\ldots$ varying from 1 to 3 mark components of the three-vectors. As usual, the sum over repeated indices is assumed. 

The first term in brackets on the right-hand side $\mathcal{S}[\mathcal{F}_{s}]$ is the source term which describes the pair creation due to the Schwinger effect. We will discuss this term in more detail below. The second term in brackets on the right-hand side of Eq.~(\ref{Boltzmann-covariant}) is the collision integral. In our work, we consider it 
in the relaxation time approximation developed first  for a relativistic plasma in Ref.~\cite{Anderson:1974}
\begin{equation}
\label{collision-integral-SRTA}
\mathcal{C}[\mathcal{F}_{s}]=-\frac{\mathcal{F}_{s}(x, P)-\mathcal{F}_{\!s,({\rm eq})}(x,P)}{\tau}.
\end{equation}
Here $\mathcal{F}_{\!s,({\rm eq})}(x, P)=\left\{\exp[(u^{\nu}P_{\nu}-\mu_{s}(x))/T(x)] \mp 1\right\}^{-1}$ is  the Bose-Einstein (sign minus) 
or Fermi-Dirac (sign plus) local equilibrium distribution function for boson or fermion particles with generally spacetime-dependent 
chemical potential $\mu_{s}(x)$ and  temperature $T(x)$ both of which should be determined self-consistently from the function $\mathcal{F}_{s}(x,P)$. Relaxation time  $\tau$ is defined by momentum relaxation of the charged particles due to their scattering mediated by the EM interaction and, therefore, could be estimated as $\tau\propto [e^{4}T {\rm ln}(1/|e|)]^{-1}$ \cite{Baym:1990,Arnold:2000,Thoma:2009a,Thoma:2009b}.

Multiplying Eq.~(\ref{Boltzmann-covariant}) by $P^{\nu}$, integrating  over momentum, and taking into account the contributions of 
particles and antiparticles with all spin projections, we obtain the following covariant energy conservation law for charged particles 
\cite{Gatoff:1987}:
\begin{equation}
\label{energy-conservation-particles}
\nabla_{\mu}T_{\rm kin}^{\mu \nu}=eF^{\nu}_{\,\,\mu}j_{\rm cond}^{\mu}+\Sigma^{\nu},
\end{equation}
where 
\begin{eqnarray}
T_{\rm kin}^{\mu\nu}&=&\sum_{s=p,a}g\int d\Gamma_{\!P} P^{\mu}P^{\nu}\mathcal{F}_{s}(x,P),\label{stress-energy-particles}\\
ej^{\mu}_{\rm cond}&=&\sum_{s=p,a}e_{s}g\int d\Gamma_{\!P} P^{\mu} \mathcal{F}_{s}(x,P),\label{cond-current-particles}\\
\Sigma^{\mu}&=&\sum_{s=p,a}g\int d\Gamma_{\!P} P^{\mu} (u^{\nu}P_{\nu}) \mathcal{S}[\mathcal{F}_{s}] \label{sigma}
\end{eqnarray}
are the stress-energy tensor  which describes the kinetic motion of the particles in the flow, the conduction electric current of moving charged particles, and  a new term accompanying the pair creation process, respectively,  and $\nabla_{\mu}$ denotes the covariant derivative. Here $g=2s+1$ is the number of spin projections and the invariant integration measure in the momentum space  is $d\Gamma_{\!P}=dP_{1}dP_{2}dP_{3}/[(2\pi)^{3}\sqrt{-{\rm det}(g_{\mu\nu})}P^{0}]$.  Note that the collision integral does not contribute to Eq.~(\ref{energy-conservation-particles}) because energy and momentum are conserved during collisions. Considering the system ``EM field+particles'' as a closed one and requiring the  conservation of the total energy and momentum, we conclude that the pair creation process leads to an additional contribution to the current known as the polarization current which is determined by the following relation:
\begin{equation}
\label{pol-current-definition}
eF^{\nu}_{\,\,\mu}j_{\rm pol}^{\mu}=\Sigma^{\nu}.
\end{equation}

Now let us focus on the case of the spatially homogeneous time-dependent electric field relevant in the study of inflationary magnetogenesis. Using the  physical momentum  of the particle $\mathbf{p}$ [so that $P^{\mu}=(\epsilon_{\mathbf{p}},\,\mathbf{p}/a)$ and $\epsilon_{\mathbf{p}}=\sqrt{m^{2}+\mathbf{p}^{2}}$], we rewrite Eq.~(\ref{Boltzmann-covariant}) for particles in the following simple form:
\begin{equation}
\label{Boltzmann-final}
\frac{\partial\mathcal{F}_{p}}{\partial t}+\mathcal{E}(t)\frac{\partial\mathcal{F}_{p}}{\partial p_{\parallel}}-H \mathbf{p}\frac{\partial\mathcal{F}_{p}}{\partial \mathbf{p}}=\mathcal{S}[\mathcal{F}_{p}]+\mathcal{C}[\mathcal{F}_{p}],
\end{equation}
where  $\mathcal{E}(t)=eE(t)$ was defined in the previous section. Here, we took into account that the created particles are uniformly distributed in coordinate space. Since the corresponding antiparticles acquire the velocity in the opposite direction, an obvious symmetry property takes place
\begin{equation}
\label{symmetry}
\mathcal{F}_{a}(t,p_{\parallel},\mathbf{p}_{\perp})=\mathcal{F}_{p}(t,-p_{\parallel},\mathbf{p}_{\perp}),
\end{equation}
where the parallel and transverse components of the momentum are determined with respect to the electric field direction. Since the distribution function for the antiparticles satisfies  Eq.~(\ref{Boltzmann-final}) with the replacements $\mathcal{E} \rightarrow -\mathcal{E}$ and $\mu \rightarrow -\mu$
we conclude that the chemical potential equals to zero, $\mu(t)\equiv 0$.  Consequently, we could omit the index $s$ in the kinetic function and consider only the distribution function for particles in what follows.

The energy density of created particles is given by the 00 component of the stress-energy tensor (\ref{stress-energy-particles})
\begin{equation}
\label{en-dens}
\rho_{c}(t)=2g\int\frac{d^{3}\mathbf{p}}{(2\pi)^{3}}\epsilon_{\mathbf{p}}\mathcal{F}(t,\mathbf{p}).
\end{equation}
 The effective local equilibrium temperature is defined by requiring that the local equilibrium distribution function  reproduces the
energy density (\ref{en-dens}), i.e.,
\begin{equation}
\rho_{c, ({\rm eq})}(t)\equiv 2g\int\frac{d^{3}\mathbf{p}}{(2\pi)^{3}}\epsilon_{\mathbf{p}}\mathcal{F}_{({\rm eq})}(t,\mathbf{p})=\rho_{c}(t).
\end{equation}
 In the ultrarelativistic limit $m\ll T$, the integral can be calculated explicitly that gives
\begin{equation}
\label{temperature-self-consistent}
T(t)=\left[\left(\frac{8}{7}\right)^{\!\!\eta}\!\frac{15}{g\pi^{2}}\,\rho_{c}(t)\right]^{1/4}=\left[\left(\frac{8}{7}\right)^{\!\!\eta}\!\frac{15}{2\pi^{4}}\int_{-\infty}^{+\infty}dp_{\parallel}\int_{0}^{+\infty}dp_{\perp}\,p_{\perp}\epsilon_{\mathbf{p}}\mathcal{F}(t,\mathbf{p})\right]^{1/4},
\end{equation}
where $\eta=0$ for bosons and $\eta=1$ for fermions.

Now let us calculate the current of created particles. Equation~(\ref{cond-current-particles})  defines the conduction part
\begin{equation}
j^{i}_{\rm cond}=\frac{2g}{a(t)}\int\frac{d^{3}\mathbf{p}}{(2\pi)^{3}}\frac{p^{i}}{\epsilon_{\mathbf{p}}}\mathcal{F}(t,\mathbf{p}).
\end{equation}
 Considering the $\nu=0$ component of Eqs.~(\ref{pol-current-definition}) and~(\ref{sigma}), we find the parallel component of the 
polarization current
\begin{equation}
j^{\parallel}_{\rm pol}=\frac{2g}{a(t)\mathcal{E}}\int\frac{d^{3}\mathbf{p}}{(2\pi)^{3}}\epsilon_{\mathbf{p}}\mathcal{S}[\mathcal{F}].
\end{equation}
 Thus, the full electric current which appears on the right-hand side of the Maxwell equation (\ref{eq-for-electric-field}) equals
\begin{equation}
\label{current-total}
\tilde{j}=a(t)(j^{\parallel}_{\rm cond}+j^{\parallel}_{\rm pol})=2g\int\frac{d^{3}\mathbf{p}}{(2\pi)^{3}}\left[\frac{p^{\parallel}}{\epsilon_{\mathbf{p}}}\mathcal{F}(t,\mathbf{p})+\epsilon_{\mathbf{p}}\frac{\mathcal{S}[\mathcal{F}]}{\mathcal{E}} \right].
\end{equation}
 Since only this current will be used, we will omit the symbol tilde in what follows.

\subsection{Source term}

The  last thing that we need to specify in order to close the system of equations is the source term in the Boltzmann equation. 
Unfortunately, it is not unambiguously fixed in the literature. Recall that the Schwinger production rate is given by the integral over 
momenta of produced particles and to obtain the information about the distribution of  created particles in momentum space is more 
difficult. This is one of the reasons why we will check our results obtained in the kinetic theory by using the hydrodynamic approach in
Sec.\ref{sec-hydro}. 

 The Schwinger source term in Minkowski spacetime was derived by using the field-theoretic methods in Refs.~\cite{Rau:1994,Schmidt:1998,Kluger:1998}. However, for the case of the expanding universe, such an expression does not exist in the literature and its derivation deserves a separate  investigation. For the purpose of the present paper, it is sufficient to use the model expression which gives the correct Schwinger pair creation rate after the integration over the momentum space and has a physically meaningful asymptotic behavior. 

 The  simplest form of the source term given in Refs.~\cite{Gatoff:1987,Kajantie:1985,Kluger:1991,Kluger:1992} assumes that particles are created with zero projection of momentum on the electric field direction
\begin{equation}
\label{source-term}
\mathcal{S}[\mathcal{F}]=(1 \pm 2\mathcal{F}(t,\mathbf{p}))|\mathcal{E}(t)| \exp\left(-\pi\frac{m^{2}+\mathbf{p}_{\perp}^{2}}{|\mathcal{E}(t)|}\right)\delta(p_{\parallel}),
\end{equation}
where the sign plus or minus corresponds to  boson or fermion particles. This expression contains the factor $(1\pm 2\mathcal{F})$ which takes into account the stimulated pair creation  for bosons and the Pauli blocking for fermions. However, the kinetic equation with the source term (\ref{source-term}) is unsuitable for numerical treatment because of the presence of the singular delta function.  Moreover, such a choice is not physically motivated because the virtual particles and antiparticles are accelerated by the electric field until their energy is sufficient to their real existence. As we will see below, the typical longitudinal momentum acquired in this process is of order $p_{\parallel}\sim \sqrt{\mathcal{E}}$. Therefore, we will consider the  source which is smooth and isotropic  in momentum space
\begin{equation}
\label{source-term-2}
\mathcal{S}[\mathcal{F}]=(1 \pm 2\mathcal{F}(t,\mathbf{p}))\sqrt{|\mathcal{E}(t)|} \exp\left(-\pi\frac{m^{2}+\mathbf{p}^{2}}{|\mathcal{E}(t)|}\right),
\end{equation}
 and will use it in numerical calculations in Sec.~\ref{sec-numerical}. 

Since the source term plays a crucial role in our kinetic approach to the Schwinger effect, let us discuss the polarization current generated by this source in more detail. For this, we omit the Bose enhancement or Fermi blocking factor  $[1\pm 2\mathcal{F}(t,\mathbf{p})]$ in expressions (\ref{source-term}) and (\ref{source-term-2}), which was added phenomenologically in order to respect the effects of the quantum statistics.
First of all, both expressions are normalized in order to reproduce the  Schwinger pair production rate in the unit volume per spin projection \cite{Schwinger:1951}
\begin{equation}
\label{gamma}
\Gamma=\int\frac{d^{3}\mathbf{p}}{(2\pi)^{3}}\, \mathcal{S}[\mathcal{F}=0]=\frac{|\mathcal{E}|^{2}}{(2\pi)^{3}}\exp\left(-\frac{\pi m^{2}}{|\mathcal{E}|}\right).
\end{equation}
In the strong field regime  $|\mathcal{E}|\gg H^{2},\ m^{2}$ realized during inflation (see the discussion in Sec.~\ref{sec-numerical}), 
the polarization current reads as
\begin{equation}
\label{j-pol-Mink}
j_{\rm pol}=\frac{2g}{\mathcal{E}}\int\frac{d^{3}\mathbf{p}}{(2\pi)^{3}}\sqrt{m^{2}+\mathbf{p}^{2}}\, \mathcal{S}[\mathcal{F}=0]\simeq g\ {\rm sign}(\mathcal{E})\frac{|\mathcal{E}|^{3/2}}{(2\pi)^{3}}
\end{equation}
for source (\ref{source-term}) and  with additional factor $4/\pi$ for source (\ref{source-term-2}).

Let us  see that this polarization current agrees well with the total Schwinger current calculated from the first 
principles by the quantum field-theoretic methods in Minkowski spacetime in Refs.~\cite{Anderson:2014a,Anderson:2014b}
\begin{equation}
\label{current-Mink}
j_{\rm tot}\simeq g\ {\rm sign}(\mathcal{E})\,\frac{\mathcal{E}^{2}}{4\pi^{3}}t,
\end{equation}
 assuming that the electric field was switched on at $t_{0}=0$.  The linear  dependence of the total current on time can be explained as follows. The electric field produces a constant number $\Gamma$ of particle-antiparticle pairs of each spin projection per unit volume and per unit time and after that accelerates the created particles. In the strong field regime, we can neglect the mass and assume that the particles almost instantly acquire velocity $v\sim 1$. Then the current density is proportional to the number of charge carriers reaching the point of observation from  $t_0=0$. Taking into account the equal contribution from antiparticles, we obtain the total current  $j_{\rm tot}=2g\,{\rm sign}(\mathcal{E})\,\Gamma t$ which coincides with Eq.~(\ref{current-Mink}) taking into account the Schwinger production rate~(\ref{gamma}). In order to compare the polarization current (\ref{j-pol-Mink}) with the total Schwinger
current (\ref{current-Mink}), we should determine the pair creation time. For this purpose, we take into account that particles' momentum grows linearly with time, $p=\mathcal{E} t$. In order to prevent the annihilation, the electric field has to separate the particle and antiparticle
by a distance of at least the  corresponding de Broglie wavelength $\lambda_{dB}$. Since we can neglect the mass in the strong field regime, we have $\Delta x\geq \lambda_{dB}=1/p$. Then, for $v \sim 1$, we obtain $t=|\Delta x|\geq 1/(|\mathcal{E}| t)$, which gives an estimate of the pair creation time $t_{\rm pc}\sim 1/\sqrt{|\mathcal{E}|}$.

Substituting  $t_{\rm pc}$ into Eq.~(\ref{current-Mink}), we find $j_{\rm tot}(t_{\rm pc})= \frac{g}{4\pi^{3}}{\rm sign}(\mathcal{E})\, |\mathcal{E}|^{3/2}\sim j_{\rm pol}$. This result reveals the physical meaning of the polarization current. It is the current of virtual particles and antiparticles before their separation.  The conduction current, in turn, is caused by real particles. Since the number of created particles grows linearly in time, the polarization current in Minkowski spacetime quickly becomes much smaller than the conduction one.

Now let us consider the case of an expanding universe. In de Sitter space, the Schwinger current created by a constant electric field in the strong field regime was calculated in Refs.~\cite{Kobayashi:2014,Bavarsad:2016} and equals
\begin{equation}
j=g\frac{\mathcal{E}^{2}}{12\pi^{3}}\frac{1}{H}.
\label{current-expanding}
\end{equation}
It is instructive to compare this expression with current~(\ref{current-Mink}) in flat spacetime. Clearly, current (\ref{current-expanding})
has a similar structure but the linear time growth  is replaced with $t_{H}\sim 1/H$ in de Sitter spacetime. It is 
easy to understand the reason for such a replacement. In an exponentially expanding background, only particles or antiparticles inside the 
Hubble horizon could reach the point of observation. Therefore, the current grows linearly in time only during the first Hubble time as the 
electric field is switched on and then remains constant. Therefore, if the characteristic pair creation time is much smaller than the Hubble 
time (this is true in the strong field regime where $t_{\rm pc}\ll t_{H}$), the situation is the same as in flat 
spacetime.  Therefore, since the source term describes the moment of the pair creation, we can use the same expression 
(\ref{source-term}) or (\ref{source-term-2}) in an expanding universe too.

\section{Hydrodynamic approach}
\label{sec-hydro}
\vspace{5mm}

The kinetic approach discussed in the previous sections is the most general way to describe the charged particles produced due to  the Schwinger effect. It gives full information about the distribution  of particles and antiparticles in the phase space allowing one to calculate any observable quantity depending on the distribution function. On the other hand, in this approach, one has to solve the partial differential equation  in spacetime coordinates and momentum which is computationally costly. Moreover, the particle distribution depends on the explicit form of the source term and the collision integral which can be modeled in different ways. Therefore, it  is useful to consider a hydrodynamic approach  that operates with the integral quantities such as number or energy densities, currents, etc. in spacetime, rather than with the distribution function in the phase space.

 Since the hydrodynamic approach operates with moments of the distribution function, let us introduce a $(k,r)$ moment
\begin{equation}
\label{moments}
\mathcal{J}^{(k,r)}_{i_{1}\ldots i_{k}}(t)= 2g\int \frac{d^{3}\mathbf{p}}{(2\pi)^{3}}\frac{p_{i_{1}}\cdots p_{i_{k}}}{\epsilon(\mathbf{p})^{r+2k-1}}\mathcal{F}(t,\mathbf{p}),
\end{equation}
where the factor 2 comes from taking into account the contributions from particles and antiparticles.  We assume that the electric field is applied along the $z$ axis. Then multiplying the Boltzmann equation (\ref{Boltzmann-final}) by  $p_{i_{1}}\cdots p_{i_{k}}/\epsilon(\mathbf{p})^{r+2k-1}$ and integrating over momentum, we obtain the following chain of equations:
\begin{multline}
\label{hydro-general}
\left[\frac{d}{dt}+\frac{1}{\tau}+(4-k-r)H\right]\mathcal{J}^{(k,r)}_{i_{1}\ldots i_{k}}+\\
+\mathcal{E}\left[(r+2k-1)\mathcal{J}^{(k+1,r)}_{i_{1}\ldots i_{k},z}-\sum_{l=1}^{k}\delta_{i_{l},z}\mathcal{J}^{(k-1,r+2)}_{i_{1}\ldots\hat{i}_{s}\ldots i_{k}}\right]
+(r+2k-1)H m^{2} \mathcal{J}^{(k,r+2)}_{i_{1}\ldots i_{k}}=\\=\mathfrak{s}^{(k,r)}_{i_{1}\ldots i_{k}}+\frac{1}{\tau}\mathcal{J}^{(k,r)}_{i_{1}\ldots i_{k},({\rm eq})},
\end{multline}
where the notation $\hat{i}_{s}$ means that the index $i_{s}$ is absent in the sequence,
\begin{equation}
\mathfrak{s}^{(k,r)}_{i_{1}\ldots i_{k}}= 2g\int \frac{d^{3}\mathbf{p}}{(2\pi)^{3}}\frac{p_{i_{1}}\cdots p_{i_{k}}}{\epsilon(\mathbf{p})^{r+2k-1}}\mathcal{S}(\mathbf{p})
\end{equation}
and
\begin{equation}
\mathcal{J}^{(k,r)}_{i_{1}\ldots i_{k},({\rm eq})}= 2g\int \frac{d^{3}\mathbf{p}}{(2\pi)^{3}}\frac{p_{i_{1}}\cdots p_{i_{k}}}{\epsilon(\mathbf{p})^{r+2k-1}}\mathcal{F}_{({\rm eq})}(\mathbf{p},T).
\end{equation}

The last term on the left-hand side of Eq.~(\ref{hydro-general}) can be neglected in the regime $\mathcal{E}\gg m^{2}$ which is the main interest for us. The effective temperature is determined  by Eq.~(\ref{temperature-self-consistent}). We
consider the first three moments (\ref{moments}) which are important for the description of the Schwinger pair production, namely,  the energy density of created particles $\mathcal{J}^{(0,0)}\equiv \rho_{c}$, their total number density $\mathcal{J}^{(0,1)}\equiv n$, and the conduction current $\mathcal{J}^{(1,0)}_{z}\equiv j_{\rm cond}$. Then the system of equations describing their evolution  is
\begin{eqnarray}
\frac{dn}{dt}+\left(3H+\frac{1}{\tau}\right)n&=&2g\Gamma +\frac{n_{({\rm eq})}}{\tau},\label{eq-concentration}\\
\frac{d\rho_{c}}{dt}+4H\rho_{c}&=&\mathcal{E}j_{\rm cond}+\mathcal{E}j_{\rm pol},\label{eq-en-dens}\\
\frac{dj_{\rm cond}}{dt}+\left(3H+\frac{1}{\tau}\right)j_{\rm cond}&=&\mathcal{E}\mathcal{J}^{(2,0)}_{\perp\perp},\label{eq-jc}
\end{eqnarray}
where $\Gamma=\int\frac{d^{3}\mathbf{p}}{(2\pi)^{3}}\mathcal{S}$
is  the number of pairs created due to  the Schwinger effect  per unit time and spin projection,
\begin{equation}
n_{({\rm eq})}=\left(\frac{3}{4}\right)^{\!\!\eta}\frac{2g\zeta(3)}{\pi^{2}}T^{3}
\end{equation}
is the equilibrium number density, and $j_{\rm pol}$ is  the polarization current. It should be noted that from the definition of the effective temperature (\ref{temperature-self-consistent}) one immediately obtains $\rho_{c,({\rm eq})}(T)\equiv \rho_{c}$, which explains why terms with the relaxation time are absent in Eq.~(\ref{eq-en-dens}). 

Obviously,  the chain of equations (\ref{hydro-general}) can be truncated only if the source term $\mathcal{S}$ does not depend on the distribution function, otherwise, $\Gamma$ and $j_{\rm pol}$ are new dynamical variables which need additional equations for their determination. Therefore, in order to proceed with the hydrodynamic approach, we have to neglect the Bose enhancement  or the Fermi blocking factor in the source term. Then the pair creation rate and polarization current can be calculated as functions of the electric field. In particular, for the first source  term~(\ref{source-term}), we get the pair creation rate (\ref{gamma}) and the polarization current (\ref{j-pol-Mink}). For  source~(\ref{source-term-2}), we obtain the same pair creation rate (\ref{gamma}) and the polarization current (\ref{j-pol-Mink}) multiplied by factor $4/\pi$.

Finally, Eq.~(\ref{eq-jc}) contains the only unknown quantity
\begin{equation}
\mathcal{J}^{(2,0)}_{\perp\perp}\equiv \mathcal{J}^{(2,0)}_{xx}+\mathcal{J}^{(2,0)}_{yy}=2g\int\frac{d^{3}\mathbf{p}}{(2\pi)^{3}}\frac{p_{\perp}^{2}}{\epsilon_{\mathbf{p}}^{3}}\mathcal{F}(t,\mathbf{p}),
\end{equation}
which has to be expressed in terms of the known variables in order to close the system of equations. We will  do this by using some phenomenological arguments. Let us assume that the particles flow can be characterized by a velocity $v$. Then, the  conduction current can be written as $j_{\rm cond}=nv$. Using Eqs.~(\ref{eq-concentration}) and (\ref{eq-jc}), we derive the  following equation for velocity $v$:
\begin{equation}
\label{eq-velocity-1}
\frac{dv}{dt}=-\frac{v}{n}\left(2g\Gamma +\frac{n_{({\rm eq})}}{\tau}\right)+\frac{\mathcal{E}}{n}\mathcal{J}^{(2,0)}_{\perp\perp}.
\end{equation}
Here, the first term on the right-hand side describes the slowing down of the particles' flow due to the relaxation processes (scattering) and  the creation of new particles which initially have zero velocity because the source term is parity even. The second term, on the contrary, is proportional to the electric field and accelerates the flow. It corresponds to the force volume density term in the Navier-Stokes equation. Assuming that the flow consists of particles with almost equal velocities and using the relativistic equation of motion for charged particles, we  obtain
\begin{equation}
\label{eq-velocity-2}
\frac{dv}{dt}=- Cv+\mathcal{E}\frac{n}{\rho_{c}}(1-v^{2}),
\end{equation}
where $C$ is the decay constant due to the processes discussed above. Comparing Eqs.~(\ref{eq-velocity-1})  and (\ref{eq-velocity-2}) and taking into account that $v=j_{\rm cond}/n$, we  find
\begin{equation}
\label{j20}
\mathcal{J}^{(2,0)}_{\perp\perp}=\frac{n^{2}-j_{\rm cond}^{2}}{\rho_{c}}.
\end{equation}

The same estimate can be done also directly from the definition of $\mathcal{J}^{(2,0)}_{\perp\perp}$
\begin{equation}
\mathcal{J}^{(2,0)}_{\perp\perp}=n\left<\frac{p_{\perp}^{2}}{\epsilon_{\mathbf{p}}^{3}}\right>\simeq n \frac{1-\left<\frac{p_{\parallel}^{2}}{\epsilon_{\mathbf{p}}^{2}}\right>}{\left<\epsilon_{\mathbf{p}}\right>}\simeq n \frac{1-\left<\frac{p_{\parallel}}{\epsilon_{\mathbf{p}}}\right>^{2}}{\left<\epsilon_{\mathbf{p}}\right>}=\frac{n^{2}-j_{\rm cond}^{2}}{\rho_{c}}, 
\end{equation}
where we took into account that, by definition, $j_{\rm cond}=n\left<\frac{p_{\parallel}}{\epsilon_{\mathbf{p}}}\right>$, $\rho_{c}=n\left<\epsilon_{\mathbf{p}}\right>$, and the angular brackets $\left<\ldots\right>$ denote the averaging with the momentum distribution function. This estimate can be used  when the particle distribution is localized in momentum space in a relatively small region far from the origin (i.e.,  when particles have almost equal nonzero momenta). In  a real situation, the electric field accelerates particles providing the displacement of their distribution from the origin; however, the characteristic width of the distribution is comparable with the displacement itself. Nevertheless, we will see below that  this approximation reproduces correctly the qualitative behavior of the system.

\section{Numerical results and discussion}
\label{sec-numerical}
\vspace{5mm}

In order to perform numerical studies, we have to specify the inflationary model. We choose the $R^{2}$ Starobinsky model
\cite{Starobinsky:1980} which is in a good agreement with the results of CMB observations \cite{Planck:2018inf} and gives the following potential for the inflaton field:
\begin{equation}
\label{Starobinsky-pot}
V(\phi)=\frac{3\mu^{2}M_{p}^{2}}{4}\left[1-\exp\left(-\sqrt{\frac{2}{3}}\frac{\phi}{M_{p}}\right)\right]^{2},
\end{equation}
where $\mu\approx 1.3\times 10^{-5}\,M_{p}$. The kinetic coupling function $f(\phi)$ is chosen  as in the Ratra model \cite{Ratra:1992} and equals
\begin{equation}
\label{ratra-function}
f(\phi)=\exp\left(\beta \frac{\phi}{M_{p}}\right),
\end{equation}
where $\beta$ is a dimensionless parameter. Previously, this model was studied in Refs.~\cite{Vilchinskii:2017,Sobol:2018} where it was shown that generated electric fields are much stronger than magnetic ones and for $\beta\gtrsim 8$ they backreact  on the Universe's expansion by slowing down the inflaton and extending the inflation stage. Moreover, using the expressions for the Schwinger current derived in the approximation of a constant electric field, it was shown that the Schwinger effect becomes important at the end of inflation when it abruptly diminishes the electric field and helps to finish the inflation stage. Therefore, it is interesting for us to extend the previous studies and  reexamine the Schwinger effect by applying a more accurate kinetic approach.

In previous sections we formulated the system of equations governing the joint evolution of the scale factor, the inflaton field, the electric field, and charged particles produced due to the Schwinger effect  given by the Friedmann equation~(\ref{eq-Friedmann}), the Klein-Gordon equation~(\ref{KGF-2}), the Maxwell equation~(\ref{eq-for-electric-field}) together with the Boltzmann equation~(\ref{Boltzmann-final}) with the collision integral (\ref{collision-integral-SRTA}) and the source term (\ref{source-term-2}). The energy density of the produced particles (\ref{en-dens}) determines  an effective temperature through relation (\ref{temperature-self-consistent}) and the electric current can be calculated from the particle distribution function by means of Eq.~(\ref{current-total}).

The hydrodynamic approach formulated in Sec.~\ref{sec-hydro} is much simpler  because it does not include the partial differential equation for the particle distribution functions and is based on a system of ordinary differential equations in a spatially homogeneous system. Apart from the standard set of the Friedmann equation~(\ref{eq-Friedmann}), the Klein-Gordon equation~(\ref{KGF-2}), and the Maxwell equation~(\ref{eq-for-electric-field}), here we have also equations for the number density (\ref{eq-concentration}), the energy density (\ref{eq-en-dens}), and the conduction current (\ref{eq-jc}) of charged particles.

In addition, we should impose the initial conditions at the beginning of inflation. Since  initially the electric field as well as charged particles 
are absent,  we set the electric field, the distribution function, the number, energy, and current 
densities to zero
\begin{equation}
\label{init-charged}
\mathcal{E}(0)=0,\quad \mathcal{F}(0,\mathbf{p})=0, \quad n(0)=0, \quad \rho_{c}(0)=0, \quad j_{\rm cond}(0)=0.
\end{equation}
The initial value for the inflaton field is chosen so that provide at least $N=50-60$ $e$-foldings of inflation, which gives
\begin{equation}
\label{init-inflaton}
\phi_{0}=\sqrt{\frac{3}{2}} M_{p}\ln \frac{4 N}{3}\simeq 5.1-5.4\,M_{p}.
\end{equation}
Then the initial value of the first derivative of the inflaton with respect to time can be found from the Friedmann and Klein-Gordon-Fock 
equations in the slow-roll approximation
\begin{equation}
\label{initial-derivative}
\dot{\phi}_{0}=-\frac{V'(\phi_{0})M_{p}}{\sqrt{3V(\phi_{0})}}=-\frac{1}{2N}\sqrt{\frac{3}{2}}\mu M_{p}.
\end{equation}

Let us  make sure that the Schwinger effect becomes important only in the strong field regime (because our equations are based on this approximation). For this purpose let us estimate the value of the electric field when the corresponding Schwinger current is comparable with other terms in the Maxwell equation (\ref{eq-for-electric-field}), in particular, with the Hubble damping term $2H\mathcal{E}$.  We use Eq.~(\ref{current-expanding}) which gives an upper bound for the induced current because a constant electric field induces a larger current than a quickly varying field of the same strength. Then we get the condition
\begin{equation}
\frac{\mathcal{E}}{H^{2}}\sim \frac{12\pi^{3}}{e_{\rm eff}^{2}}\gg 1.
\end{equation}
Thus, the kinetic or hydrodynamic equations describing the Schwinger effect are supposed to work well in the strong  field regime.

It is important to note that the mass of charge carriers is also time dependent during inflation and  may significantly differ from 
its zero-temperature value because the Higgs field acquires a large mean value due to quantum fluctuations \cite{Shtanov:1995}
\begin{equation}
h\simeq \left(\frac{3}{\lambda\pi^{2}}\right)^{1/4}H,
\end{equation}
where $\lambda\approx 0.26$ is the Higgs self-coupling constant.  Thus, the mass of 
particles is given by
\begin{equation}
\label{mass}
m(t)=\frac{y}{\sqrt{2}}h(t)=\frac{y}{\sqrt{2}}\left(\frac{3}{\lambda\pi^{2}}\right)^{1/4}H(t)\sim y H(t),
\end{equation}
where $y$ is the Yukawa constant which equals $y_{e}=3\times 10^{-6}$ for the electron. Since for all Standard Model particles the Yukava constants are $y\lesssim 1$,  the inequality $|\mathcal{E}|\gg m^{2}$ is automatically satisfied if $|\mathcal{E}|\gg H^{2}$.

Before presenting the numerical results, we would like to discuss the relaxation time which  enters the collision integral in  the Boltzmann equation (\ref{Boltzmann-final}). It is determined by the electromagnetic interaction between charged particles and, therefore,  a power counting implies that it is proportional to $\tau\propto (e_{\rm eff}^{4}T)^{-1}$. Additional logarithmic corrections come from the fact that, not only are the large angle scatterings important but also a large number of subsequent small angle scatterings contribute at the same order \cite{Baym:1990,Arnold:2000,Thoma:2009a,Thoma:2009b}. Therefore, the final parametric dependence on the coupling constant  is
\begin{equation}
\label{tau-param}
\tau=\frac{c_{0}(4\pi)^{2}}{Te_{\rm eff}^{4}\ln|e_{\rm eff}|^{-1}},
\end{equation}
where the dimensionless constant $c_{0}$  is order unity and has different expressions in the literature depending on the 
calculation method~\cite{Baym:1990,Arnold:2000,Thoma:2009a,Thoma:2009b}. Since this value depends on the statistics of charge carriers, on the number of flavors, etc., we use the value $c_{0}=1$ in our 
numerical analysis and only at the end of this section do we study the dependence of the results on the value of $c_{0}$.

\begin{figure}[ht]
		\centering
		\includegraphics[width=0.45\textwidth]{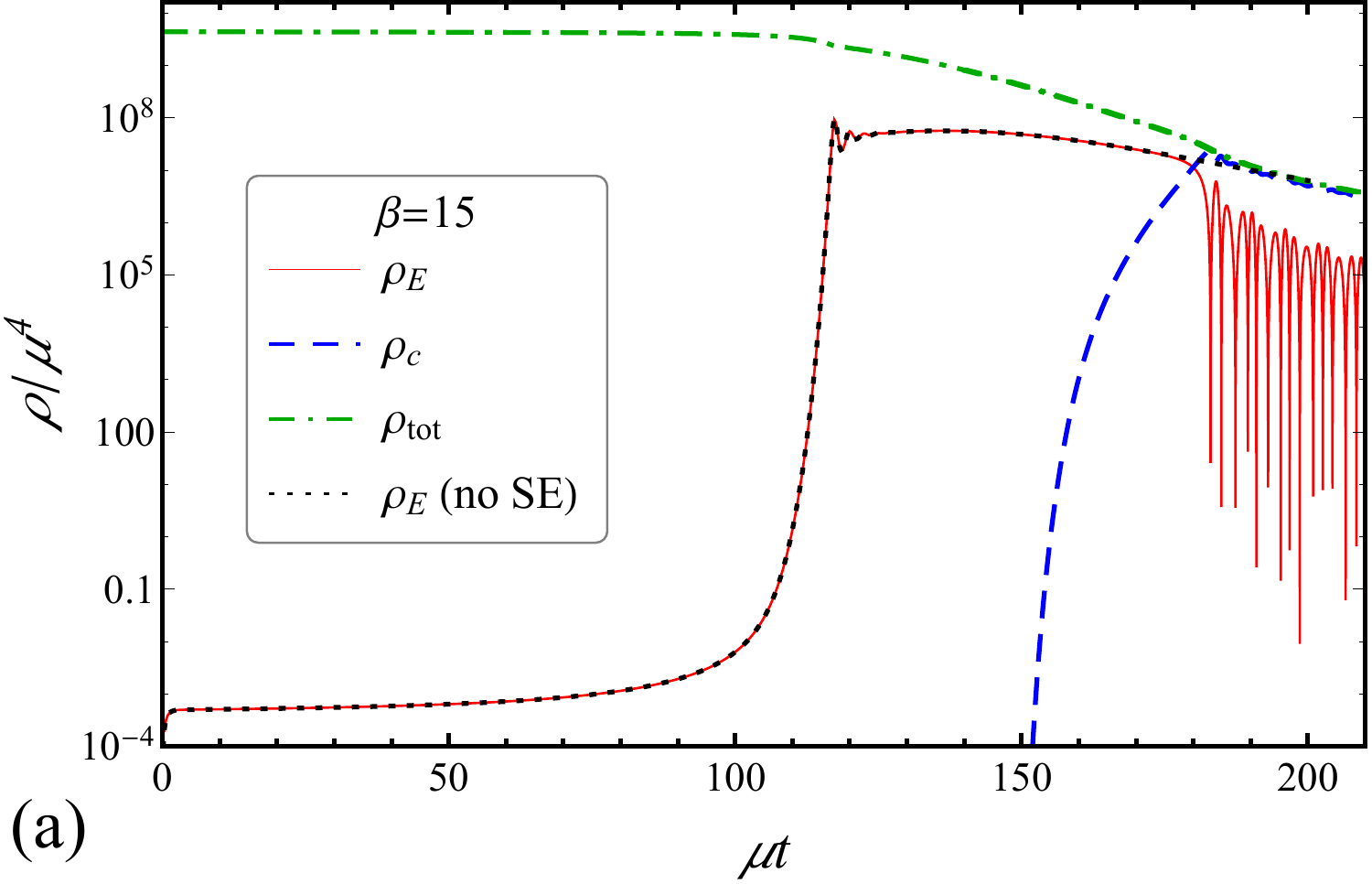}
		\includegraphics[width=0.45\textwidth]{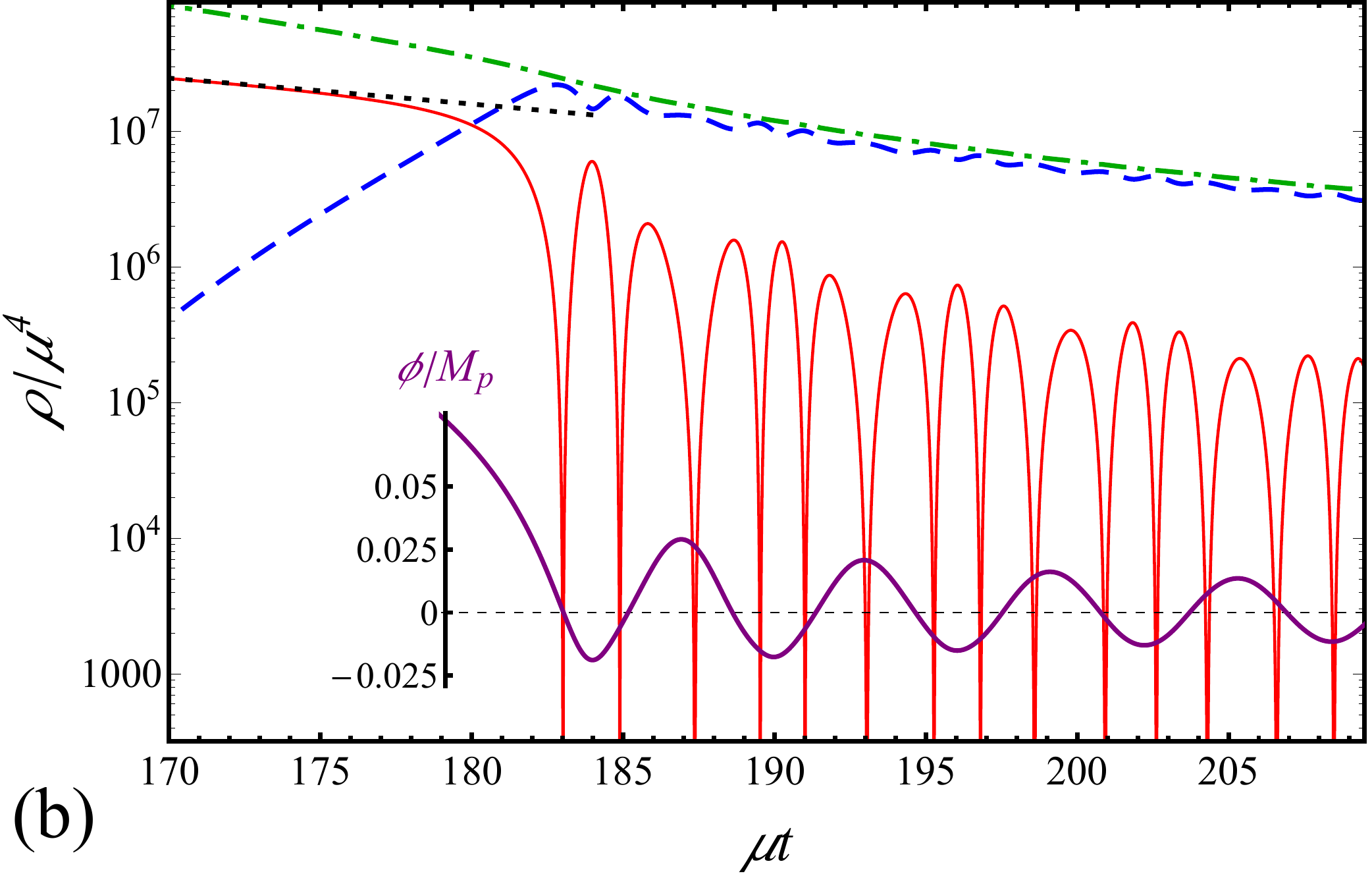}
		\caption{(a) The time dependence of the electric energy density (red solid line), the energy density of charged fermions produced due to the Schwinger effect (blue dashed line), the total energy density (green dashed-dotted line) calculated within the kinetic framework for the coupling parameter $\beta=15$. The black dotted line shows the electric energy density in the absence of the Schwinger effect. Panel (b) shows the same quantities during the final part of inflation and preheating. The inset shows the time dependence of the inflaton field. \label{fig-1-b15}}
\end{figure}

First, we study the dynamics of the Schwinger effect in  a strong electric field. For this purpose, we choose the large value of the coupling parameter $\beta=15$ for which the backreaction of the generated electric field  is relevant. Figure~\ref{fig-1-b15} shows the time dependence of the electric energy density (red solid line) together with the energy density of charged fermions produced due to the Schwinger effect (blue dashed line) calculated in the kinetic approach. The total energy density is shown by the green dashed-dotted line. For comparison, we also show the electric energy density in the absence of the Schwinger effect by the black dotted line. 

At the beginning of inflation, the electric energy density acquires  a value $\rho_{E}\sim H^{4}$ which corresponds to the characteristic amplitude of vacuum EM fluctuations. This happens because the inflaton field slowly rolls along the slope of its potential and the coupling function changes also slowly. As a result, the term $2\frac{\dot{f}}{f}\mathcal{E}$ cannot exceed the Hubble damping term $2H\mathcal{E}$ in Eq.~(\ref{eq-for-electric-field}) and the generation is absent. Close to the end of inflation, however, the situation changes  as the inflaton field rolls faster and the electromagnetic modes outside the horizon undergo significant amplification which is manifested in a steep growth of the red curve in Fig.~\ref{fig-1-b15}(a). The value of the electric energy density quickly approaches that of the inflaton $\rho_{\rm inf}$ and the backreaction comes into play when $\rho_{E}\sim \epsilon \rho_{\rm inf}$ \cite{Sobol:2018}, where  $\epsilon=(M_{p}^{2}/2)(V'/V)^{2}$ is the slow-roll parameter. This abruptly stops the growth of the electric energy density and  it remains almost constant until the very end of inflation.

Figure~\ref{fig-1-b15}(b) shows the final part of inflation and preheating in more detail.  
When the inflaton rolls down close to the minimum of its potential at $\phi=0$ (see the inset), the coupling function $f$ is of order unity and the effective charge is not suppressed any more.  Therefore, the Schwinger effect comes into play and the electric energy density starts to decrease. First of all, we  find that taking into account the dynamics of the produced particles reveals a qualitatively new effect of electric field oscillations.  At second, the energy density of the produced particles quickly grows and becomes a dominating part of the total energy density  leading to the Universe reheating. Therefore, the Schwinger reheating should be considered as an important scenario complementary to the usual ones with the inflaton perturbative decay and parametric resonance  \cite{Kofman:1997,Bassett:2006,Allahverdi:2010,Frolov:2010,Amin:2015}.

\begin{figure}[ht]
	\centering
	\includegraphics[width=0.5\textwidth]{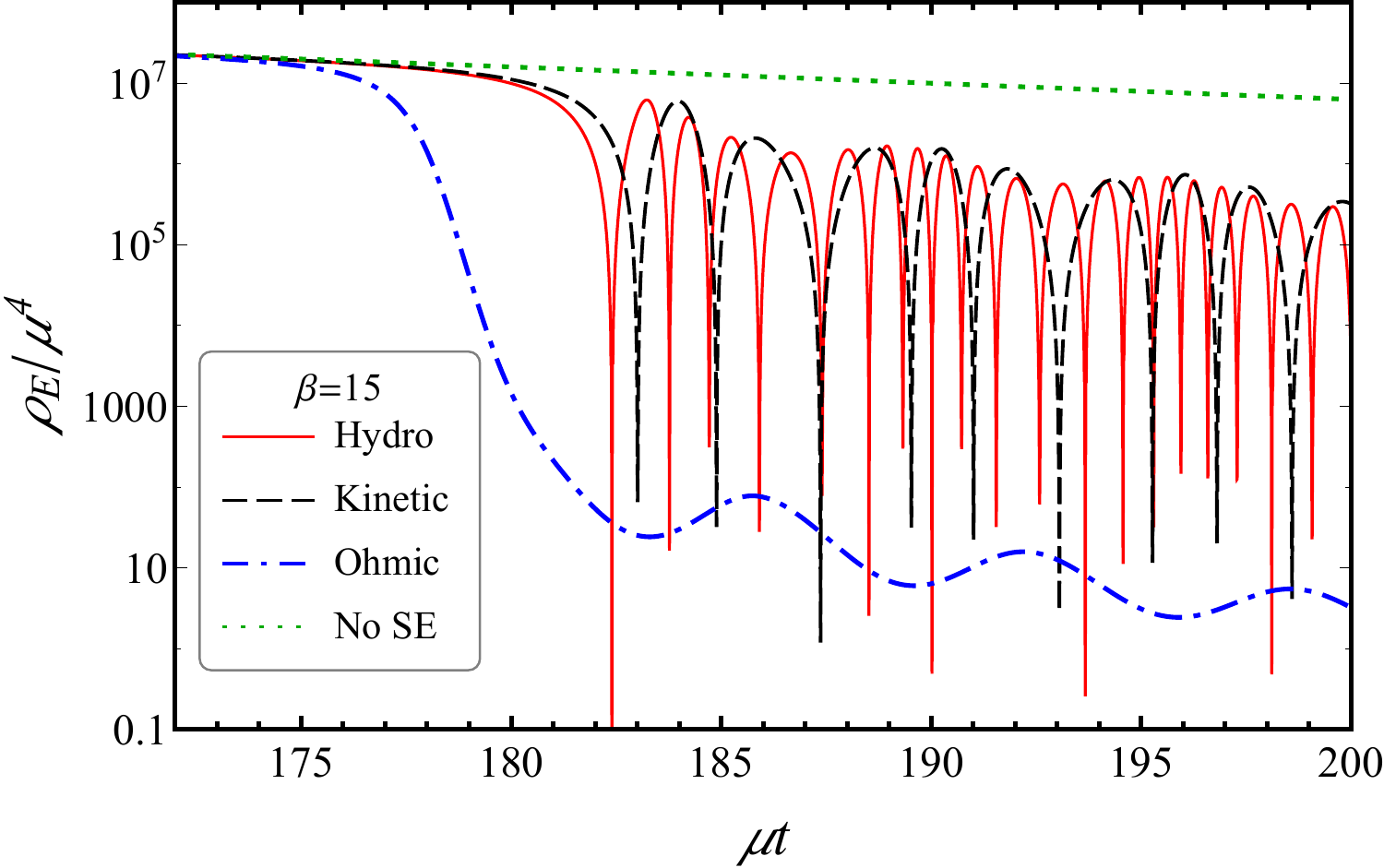}
	\caption{The time dependence of the electric energy density calculated in the hydrodynamic framework (red solid line), in the kinetic approach (black dashed line), and in the assumption that the Schwinger current  is given by Eq.~(\ref{current-expanding}) (blue dashed-dotted line). The electric energy density in the absence of the Schwinger effect is shown by the green dotted line. \label{fig-2-compare}}
\end{figure}

Let us compare the results obtained in different approaches. In Fig.~\ref{fig-2-compare} we plot the time dependence of the electric energy density calculated in the hydrodynamic framework (red solid line), in the kinetic approach (black dashed line), and in the assumption that the Schwinger current  is given by (\ref{current-expanding}) (blue dashed-dotted line). For comparison, the electric energy density in the absence of the Schwinger effect is shown by the green dotted line. We would like to note that the kinetic and the hydrodynamic approaches give qualitatively similar results with fast oscillations of the electric field. This is in contrast to the case of  Ohm's current (\ref{current-expanding}) where the electric field is abruptly damped without oscillations. However, there are some quantitative differences between the results  of the kinetic and hydrodynamic approaches. In the latter case, the Schwinger effect turns on a bit earlier and the frequency of oscillations is higher. This can be explained by the fact that  the source term (\ref{source-term}) in the Boltzmann equation contains  for fermions the Pauli blocking factor which makes the production of particles in partially filled states unfavorable.  Since the blocking factor is not included in the hydrodynamic approach, the current grows faster.  We checked that the situation is opposite for the production of charged bosons because the Bose enhancement factor makes the pair creation strongly favorable. As a result the hydrodynamic approach underestimates the Schwinger current in this case.

 The oscillatory behavior can be readily understood from the analysis of the hydrodynamical system of equations, in particular, the 
Maxwell equation~(\ref{eq-for-electric-field}) and Eq.~(\ref{eq-jc}) for the conduction current. Taking into account that  the 
period of oscillations is much less than the Hubble time $t_{H}\sim H^{-1}$, we keep only  quickly varying terms that give
\begin{equation}
\label{system-ballistic}
\frac{d\mathcal{E}}{dt}\approx -e_{\rm eff}^{2} j_{\rm cond}, \quad \frac{dj_{\rm cond}}{dt}\approx \mathcal{E} \frac{n_{({\rm eq})}^{2}}{\rho_{c, ({\rm eq})}},
\end{equation}
where  we used the equilibrium expression for $\mathcal{J}^{2,0}_{\perp\perp}$  in the second equation. Obviously, this system 
 of linear ordinary differential equations with constant coefficients
admits  an oscillatory solution for the electric field with the period
\begin{equation}
\label{osc-period}
t_{\rm osc}=2\pi \left(\frac{\rho_{c, ({\rm eq})}}{e_{\rm eff}^{2}n_{({\rm eq})}^{2}}\right)^{1/2}\sim \frac{1}{e_{\rm eff} T}.
\end{equation}
Physically, it is clear that, for $t_{\rm osc}\ll \tau,\, t_{H}$, particles on the timescale $t_{\rm osc}$ move in a time-varying 
electric field without scatterings on each other and their momenta mostly change by the electric field rather than due to the Hubble damping.
Therefore, a ballistic regime takes place and the current is retarded compared to the electric field due to the inertia of charge 
carriers. On the contrary, if the particles momenta are quickly equilibrated due to scatterings ($\tau\ll t_{\rm osc}$) or strongly damped due to the cosmological redshift ($t_{H}\ll t_{\rm osc}$), the current is determined only by the actual value of the electric field, i.e., it has the Ohmic form. This is why the phase shift between the electric field and current is absent and they monotonously decrease without oscillations (see the blue dashed-dotted line in Fig.~\ref{fig-2-compare}).

Figures~\ref{fig-1-b15} and \ref{fig-2-compare} show that the amplitude and frequency of the electric energy density oscillations are 
modulated in time.  These modulations are connected with the inflaton oscillations in the minimum of its potential as is shown in 
the inset of Fig.~\ref{fig-1-b15} (b). In fact, when the inflaton field decreases, the term $2\frac{\dot{f}}{f}\mathcal{E}$ in
Eq.~(\ref{eq-for-electric-field}) has  the negative sign and leads to an amplification of the electric field. On contrary, when the 
inflaton field grows, a damping of the electric field is observed. The period of oscillations estimated in Eq.~(\ref{osc-period}) is proportional 
to the coupling function $f$. As a result, oscillations are more frequent when the inflaton is in the minimum  of its potential and 
more sparse when it is in the maximum. These features are clearly seen  in Fig.~\ref{fig-1-b15} (b).

\begin{figure}[ht]
	\centering
	\includegraphics[width=0.5\textwidth]{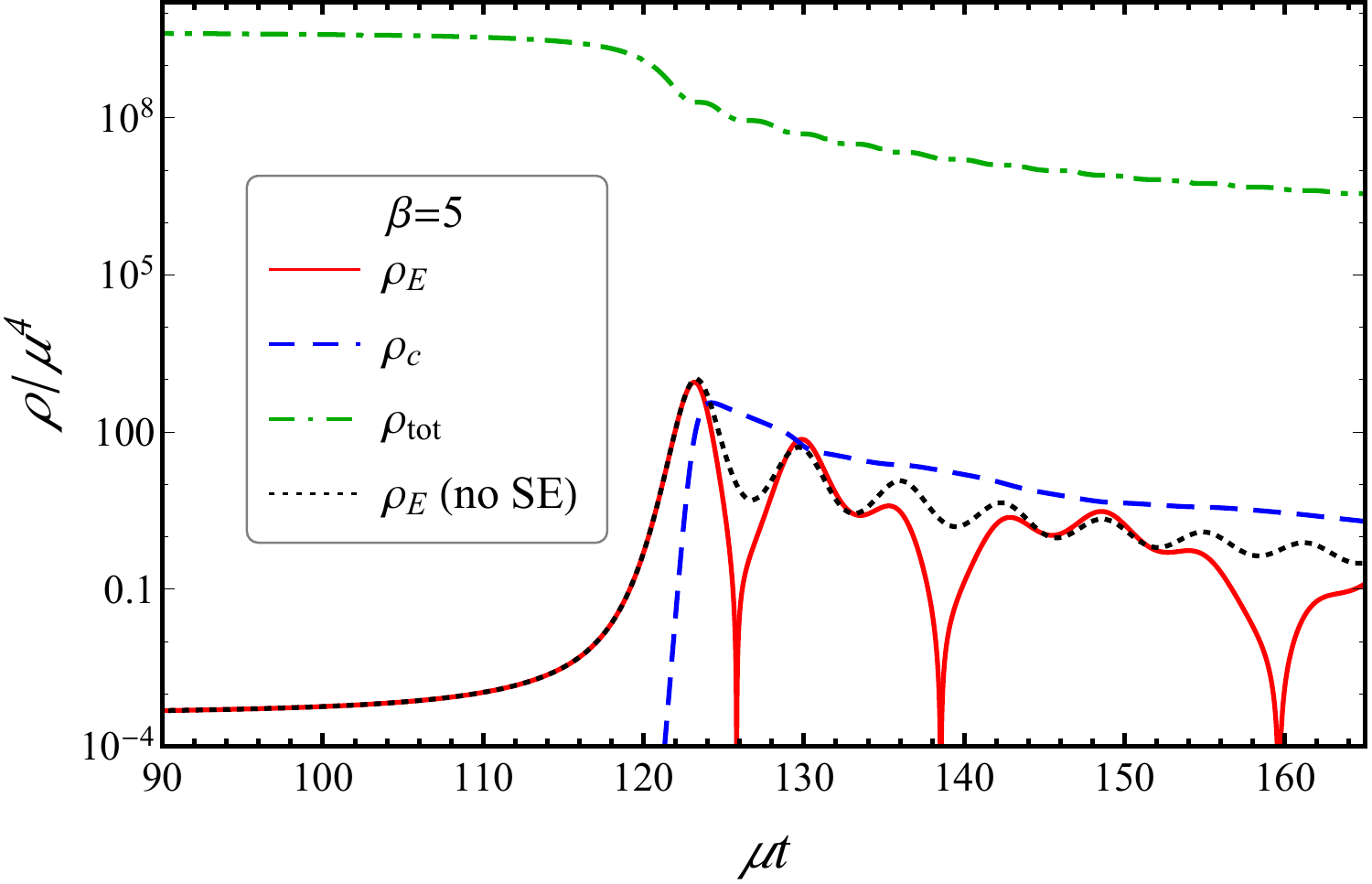}
	\caption{The time dependence of the electric energy density (red solid line), the energy density of charged fermions produced due to the Schwinger effect (blue dashed line), the total energy density (green dashed-dotted line) calculated within the kinetic framework for the coupling parameter $\beta=5$. The black dotted line shows the electric energy density in the absence of the Schwinger effect. \label{fig-3-b5}}
\end{figure}

Now let us consider the case when the electric field does not cause the backreaction on the inflaton dynamics. Figure~\ref{fig-3-b5} shows the time dependence of the energy densities for the coupling parameter $\beta=5$. When the Schwinger effect comes into play at the end of inflation, the energy density of generated charged particles reaches the value comparable to that of the electric energy density which is now much less than the inflaton contribution. The corresponding temperature is at least 1 order of magnitude less than in the previous case.  Consequently, as expected, for weak generated electric fields, the Schwinger reheating scenario is not important. In this case, the electric field also exhibits the oscillatory behavior, although these oscillations have a much bigger period. This is connected with a much lower temperature; see Eq.~(\ref{osc-period}).

\begin{figure}[ht]
	\centering
	\includegraphics[width=0.5\textwidth]{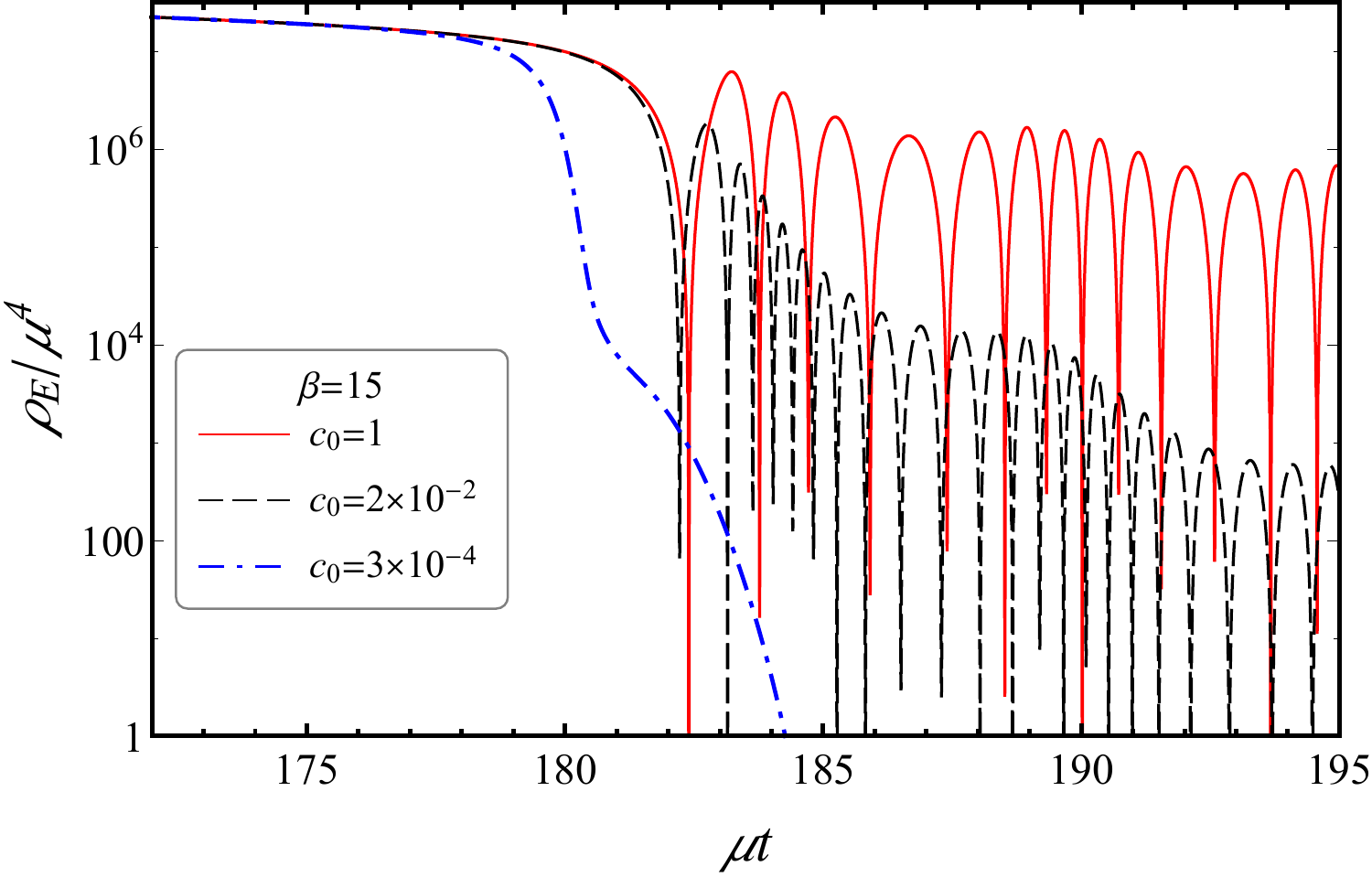}
	\caption{The time dependence of the electric energy density for different values of the parameter $c_{0}$ in Eq.~(\ref{tau-param}) determining the relaxation time: $c_{0}=1$ (red solid line), $c_{0}=2\times 10^{-2}$ (black dashed line), and $c_{0}=3\times 10^{-4}$ (blue dashed-dotted line).  \label{fig-4-tau}}
\end{figure}

Finally, we would like to study  the dependence of the results on the relaxation time. For this, we show in Fig.~\ref{fig-4-tau} the time dependence of the electric energy density calculated in the hydrodynamical approach for different values of the relaxation time parametrized by the constant $c_{0}$; see Eq.~(\ref{tau-param}). The relaxation time  $\tau$ determines the decay rate of the  amplitude of oscillations in addition to the Hubble damping. If  $\tau$ is determined by the EM processes, then $c_{0} \sim 1$, $\tau \gg t_{\rm osc}$, and the oscillatory behavior is clearly observed (red solid line). However,  if the relaxation is faster for some reason and $c_{0}\ll e_{\rm eff}^{3/2}\ll 1$,  then $\tau \ll t_{\rm osc}$ and a fast damping of the electric field occurs (see the blue dashed-dotted line). This happens because the charge carriers undergo a large number of collisions which make their movement chaotic and remove all memory effects. As a result, the drift of charged particles is determined only by the actual value of the electric field and the current has the Ohmic form.

\section{Conclusion}
\label{sec-concl}
\vspace{5mm}

In this work we  applied the kinetic and hydrodynamic approaches to the study of the Schwinger effect during inflation taking into account the dynamics of produced charged particles. To the best of our knowledge, such a self-consistent kinetic consideration of the Schwinger effect, the dynamics of produced charged particles, and the evolution of the electric and inflaton fields in the early Universe was not done in the literature before.  In order to generate electromagnetic fields during inflation, we considered the kinetic coupling of the EM field to the inflaton field in the Starobinsky model. For the coupling function $f(\phi)$ decreasing in time, the generated electric field is much stronger than the magnetic one and makes this model attractive for studying the role of the Schwinger effect in the inflationary magnetogenesis.

 In order to take into account the dynamics of created particles due to the Schwinger effect, we applied the Boltzmann kinetic equation with the source term responsible for the creation of charged particles and the collision integral describing their thermalization due to scatterings. We used the model expression for the source term previously used in the description of the Schwinger effect in heavy-ion collisions and laser beams, which reproduces the correct pair creation rate per unit time and takes into account the stimulated pair production for bosons and  the Pauli blocking for fermions.  The relaxation time approximation is used for the collision integral with $\tau\propto (e_{\rm eff}^{4} \ln |e_{\rm eff}|^{-1} T)^{-1}$ where temperature $T$ is self-consistently determined from the energy density of charged particles.  In addition, we considered also the hydrodynamic approach which operates with the number, energy, and current densities. Although it does not account for the effects of quantum statistics in the pair creation process, it is much simpler, computationally less costly, and reproduces the correct qualitative behavior of the electric field.

The key factor from the viewpoint of the evolution of EM fields is the expression for the electric current. Usually, it is taken  in 
the Ohmic form, whereas we computed it dynamically in the applied kinetic and hydrodynamic approaches. We showed that the qualitative time 
behavior of the electric field strongly depends on the interrelation between the Hubble time $t_{H}$, the relaxation time $\tau$, and the 
oscillation time $t_{\rm osc}\sim (e_{\rm eff} T)^{-1}$.
 For $t_{\rm osc}\ll t_{H},\,\tau$, the ballistic regime is realized when the electric field and current oscillate. The retardation 
of the current compared to the electric field can be explained by the inertial properties of charge carriers. On the other hand, if $t_{H}\ll \tau,\, t_{\rm osc}$ or
$\tau\ll t_{H},\, t_{\rm osc}$, then oscillations are absent and the electric field decreases monotonously with time. Remarkably, for large values of $\beta$, in addition to the oscillations, the electric energy density on average remains 4--5 orders of magnitude larger than in the case of the Ohmic electric current. This might be prospective for magnetogenesis, because the electric field can source the magnetic one through the Faraday law \cite{Kobayashi:2019}.

We found that the Schwinger process effectively produces the charged particles only close to the end of inflation when the effective charge $e_{\rm eff}=e/f$ is not suppressed by a large value of the coupling function. If the electric field is strong enough to cause the backreaction on the inflaton evolution, the energy density of charged particles produced due to the Schwinger effect becomes comparable with that of the inflaton  field. This  shows that the Schwinger effect can reheat the Universe even before the stage of fast oscillations of the inflaton.

In our work, we used the model expressions for the Schwinger source term and the collision integral in  the relaxation time 
approximation. However, it would be  interesting to perform calculations by using more realistic expressions and check how this will 
affect the oscillation behavior of the electric field and current found in this paper.
This need is especially acute for the source term describing the pair production. However, the corresponding expression derived from the first principles  is absent in the case of an expanding universe. This problem is very interesting on its own and certainly deserves a separate investigation. 

Here we numerically investigated only the simplest case of one type of charge carrier created by the Schwinger effect. It would be interesting to consider the production of all types of charged particles in the Standard Model (as well as photons, as discussed in Refs.~\cite{Akhmedov:2014,Akhmedov:2015}) and to take into account the running of the EM coupling constant $\alpha_{\rm EM}$ at high temperatures. The annihilation of the charged particles into photons might be also very important for the evolution of the primordial plasma. We plan to address these issues elsewhere.

\begin{acknowledgments}
	
	This work was supported partially by the Ukrainian State Foundation for Fundamental Research.	
	The work of O.~O.~S. was supported by the ERC-AdG-2015 Grant No. 694896 and by the Swiss National Science Foundation Grant No. 200020B\_182864.	
	The work of S.~I.~V. was supported by the Swiss National Science Foundation Grant No. SCOPE IZSEZ0-186551 and by the German Academic Exchange Service (DAAD), Grant No. 57387479. 
	S.~I.~V. and O.~O.~S. are grateful to Professor Mikhail Shaposhnikov for his kind hospitality at the Institute of Physics, \'{E}cole Polytechnique F\'{e}d\'{e}rale de Lausanne, Switzerland, where the part of this work was done.
	
\end{acknowledgments}

\end{document}